\documentclass[12pt, a4paper]{article}

\usepackage{Sweave}

\usepackage{amsmath}
\usepackage{amssymb}
\usepackage{amsfonts}
\usepackage{amsthm}
\usepackage{latexsym}
\usepackage{color}
\usepackage{epsfig}
\usepackage{natbib}
\usepackage{hyperref}

\let\proglang=\textsf
\newcommand{\pkg}[1]{{\fontseries{b}\selectfont #1}}
\let\code=\texttt

\newcommand{\blanco}[1]{ }

\newcommand{\Fm}{{\mathbb F}_m}
\newcommand{\Gn}{{\mathbb G}_n}
\def\roc{R}
\def\roce{\tilde{R}_{m, n}}
\def\rocl{\hat{R}_{m, n}}
\def\rocls{\hat{R}_{m, n}^*}
\def\rocn{\bar{R}_{m, n}}
\def\rock{\underline{\hat R}_{m, n}}

\def\dd{\, \mathrm d}                               

\def\lest{{\hat \varphi}_n}                         
\newcommand{\Hol}[3]{\HH^{#1,#2}(#3)}               

\def\toD{\to_{d}}

\def\toP{\to_{\rm p}}

\def\HH{{\cal H}}

\def\bea{\begin{eqnarray*}}
\def\eea{\end{eqnarray*}}
\def\be{\begin{equation}}
\def\ee{\end{equation}}
\def\bean{\begin{eqnarray}}
\def\eean{\end{eqnarray}}
\def\barr{\begin{array}}
\def\earr{\end{array}}
\def\bdes{\begin{description}}
\def\edes{\end{description}}
\def\bi{\begin{itemize}}
\def\ei{\end{itemize}}

\def\Bl{\Bigl}
\def\Br{\Bigr}

\def\op{o_{\rm p}}
\def\opas{o_{\rm as}}
\def\Op{O_{\rm p}}

\def\R{\mathbb{R}}

\def\ed{\end{document}}

\newtheorem{theorem}{Theorem}[section]
\newtheorem{lemma}[theorem]{Lemma}
\newtheorem{corollary}[theorem]{Corollary}

\newenvironment{Theorem}{\begin{theorem}\sl}{\end{theorem}}

\begin{document}

\title{A smooth ROC curve estimator based on log-concave density estimates}

\author{
    Kaspar Rufibach \\
    Division of Biostatistics\\
    Institute of Social and Preventive Medicine\\
    University of Zurich\\
    kaspar.rufibach@ifspm.uzh.ch\\
}

\date{\today}

\maketitle

\begin{abstract}
We introduce a new smooth estimator of the ROC curve based on log-concave density estimates of the constituent distributions. We show that our estimate is asymptotically equivalent to the empirical ROC curve if the underlying densities are in fact log-concave. In addition, we empirically show that our proposed estimator exhibits an efficiency gain for finite sample sizes with respect to the standard empirical estimate in various scenarios and that it is only slightly less efficient, if at all, compared to the fully parametric binormal estimate in case the underlying distributions are normal. The estimator is also quite robust against modest deviations from the log-concavity assumption.
We show that bootstrap confidence intervals for the value of the ROC curve at a fixed false positive fraction based on the new estimate are on average shorter compared to the approach by \cite{zhou_05}, while maintaining coverage probability.
Computation of our proposed estimate uses the \proglang{R} package \pkg{logcondens} that implements univariate log-concave density estimation and can be done very efficiently using only one line of code. These obtained results lead us to advocate our estimate for a wide range of scenarios.
\end{abstract}


\section{Introduction} \label{intro}
The receiver operating characteristic (ROC) curve is a common way of assessing the diagnostic accuracy of a diagnostic test with continuous outcome $Z$ that predicts presence or absence of a binary trait, typically a disease. The ROC curve is defined as the plot of the sensitivity (the true positive fraction)
against one minus specificity (the false positive fraction) across all possible choices of threshold values $\theta$. Thus, the ROC curve
displays the range of possible trade-offs between the true and false positive rates.
To fix ideas, let $F$ and $G$ denote distribution functions of the test result $X$ and $Y$ of a non-diseased and a diseased subject, respectively.
If $F$ and $G$ are absolutely continuous, the ROC curve of the test can be expressed as
\bean
    \roc(t; F, G) &=& 1 - G(F ^ {-1}(1 - t)) \label{def: roc}
\eean where $F^{-1}$ is the quantile function of $F$, and $t \in (0, 1)$ is the false positive fraction corresponding to a cut-off 
for positivity of the test.

We suppose that a sample $X_1, \ldots, X_m \sim F$ from the non-diseased population and a sample $Y_1, \ldots, Y_n \sim G$ from the diseased population
are available. We further assume that all these observations are mutually independent and denote the empirical distribution functions based on the samples by
$\Fm$ and $\Gn$, respectively.
The empirical quantile function for a sample $X_1, \ldots, X_m$ is defined as $\mathbb{F}^{-1}(p) = X_i$ if $\mathbb{F}(X_{i-1}) < p \le \mathbb{F}(X_{i})$
for any $p \in [0, 1]$ and by setting $X_0 = -\infty, X_{n+1} = \infty$.
In the absence of any further information about $F$ and $G$, plugging in these empirical distribution functions in \eqref{def: roc}
yields the nonparametric maximum likelihood estimator $\roce(t) = \roc(t; \Fm, \Gn)$ of $\roc(t; F, G)$, an increasing step function 
that simply consists of plotting the empirical proportions $\#\{Y_j > t\} / n$ vs. $\#\{X_i > t\}/m$ for varying $t$.
Strong consistency and strong approximation properties of $\roce$ are provided in \cite{hsieh_96}.
Being a fully nonparametric estimator, $\roce$ displays the data pattern well but may suffer from substantial variability and, as a step
function, is not smooth.
Moreover, due to the rugged form, an empirical ROC curve can have the same true positive fraction corresponding to a range of false positive fractions.
This is inconvenient when one is interested in finding a particular false positive fraction at a specified true positive fraction.

In applications, it seems sensible to assume that the true underlying ROC curve $R$ is in fact smooth. Thus we expect an estimator to
perform better if some smoothness is imposed \citep[Section 1]{lloyd_99}. The most obvious way of doing so
is to assume a parametric model, estimate the distribution functions $F$ and $G$ on the basis of the samples of healthy and diseased subjects,
and compute the corresponding ROC curve.
Assuming a normal distribution both in the control and in the cases group is the most prominent parametric model \citep[Example 5.3]{pepe_03}.

Among the frequentist approaches that are between the two ``extreme cases'', the entirely nonparametric and the parametric ROC curve estimate, are
semiparametric models. Such models do not assume a parametric form for the constituent distributions but rather for the ROC curve directly \citep{cai_04}.
The most prominent semiparametric model for the ROC curve stipulates the existence of a function $h$ such that $h(X) \sim N(0, 1)$ and
$h(Y) \sim N(\mu, \sigma^2)$. The binormal ROC curve is then given by
\bean
    \rocn(t) &=& \Phi(a + b \Phi^{-1}(t)) \label{def: binorm roc}
\eean where $a$ and $b$ are to be estimated and $\Phi$ is the cumulative distribution (CDF) of a standard normal distribution.
\cite{hsieh_96} propose a generalized least squares estimate whereas \cite{cai_04} and \cite{zhou_08} discuss
profile likelihood methods to estimate $a$ and $b$ in the general setup \eqref{def: binorm roc}.


With the purpose of defining a smooth, though more flexible than parametric, ROC curve estimate \cite{lloyd_98_jasa} proposes to plug in kernel estimates
of $F$ and $G$ into \eqref{def: roc}. As is common for all kernel estimates, a kernel and, more importantly, a bandwidth have to be chosen.
In the ROC curve estimation problem this issue is even more pronounced, since
both the ordinate and coordinate of the estimated ROC curve are random and it is thus unclear which variation to minimize in mean square.
\cite{lloyd_98_jasa} suggests to circumvent the bandwidth choice problem by a rather involved graphical examination of bias and variation from bootstrap samples.
That this kernel ROC estimate is more efficient than the empirical is shown in the subsequent paper \cite{lloyd_99}.
However, as discussed in the bandwidth comparison study performed by \cite{zhou_harez_02}, Lloyd's estimate may not be practical for routine use in
medical studies, although it has good theoretical properties. Its main drawbacks is that bandwidth selection is
ad-hoc and difficult. \cite{hall_03} take a somewhat different approach as they choose the optimal bandwidth not separately for $F$ and $G$,
but they derive optimal bandwidths via minimizing the mean integrated squared error (MISE) for the estimation of $R$ directly. They show via simulations that
their estimator improves on existing approaches in terms of MISE.

Another approach to smooth estimation is taken in \cite{qiu_01} who combine
kernel estimation of $G$ with a quantile estimator due to \cite{harrell_82}.
The latter has been shown to perform better than the usual empirical quantile
estimator in many scenarios. \cite{du_09} propose a monotone spline approach
to ensure monotonicity and transformation invariance of the estimated ROC
curve. However, here one has to select a smoothing parameter and \cite{du_09}
propose to use cross validation to guide that choice.

Further papers that deal with semiparametric approaches to estimate a ROC curve are \cite{lloyd_02}, \cite{peng_04}, and \cite{wan_07}.
However, it seems safe to say that diagnostic tests in applications are to a large extent assessed by using the empirical ROC curve estimate $\roce$
\citep[Section 1]{zhou_harez_02} and to a lesser extent the binormal estimate, despite the well-known deficiencies of the latter.
Presumably, this has to be attributed to lack of easy accessible software that implements alternative methods.

In Section~\ref{logcon} we briefly summarize some facts on log-concave density estimation. Our new estimator is introduced in Section~\ref{newROC} and some theoretical results are provided in Section~\ref{main}. An illustration of the proposed ROC curve estimate using a real-world dataset is discussed in Section~\ref{sec: illustrate roc} and its performance in simulations is assessed in Section~\ref{simulations}. Bootstrap confidence intervals for the value of the ROC curve at a fixed false positive fraction $t$ are introduced in Section~\ref{confints} and Section~\ref{misspec} discusses the features of the new estimator under misspecification. Finally, in Section~\ref{conclusions} we draw some conclusions. Additional results on log-concave density estimation, on estimation of AUC, on simulations and under misspecification as well as proofs are postponed to the appendix.

\section{Log-concave density estimation} \label{logcon}
In Section~\ref{newROC} we propose an alternative ROC curve estimator in the spirit of \cite{lloyd_98_jasa}.
Namely, we also model the constituent distributions $F$ and $G$ nonparametrically. However, we do not suggest to use kernels, but the log-concave density
estimate initially introduced in \cite{walther_02}, \cite{rufibach_06}, \cite{pal_07}, and \cite{rufibach_07}.
The features of the new ROC curve estimator are discussed in Section~\ref{newROC}.
More details on recent research on log-concave density estimation is provided in the appendix.

Rate of uniform convergence for the univariate log-concave maximum likelihood estimate was derived in \cite{duembgen_09} and its computation is described in \cite{rufibach_07}, \cite{duembgen_07},
and \cite{duembgen_logcon10}. The corresponding software is available from CRAN as the \code{R} package \pkg{logcondens} \citep{logcondens}. The multivariate case has been treated by \cite{cule_08} and the extension to discrete observations in \cite{balabdaoui_11}. For a recent review of all these developments we refer to \cite{walther_08}.

To fix ideas, let $p$ be a probability density on $\R$. We call $p$ log-concave if it may be written as $p(x) = \exp \varphi(x)$ for some
concave function $\varphi: \R \to [-\infty, \infty)$ and $x \in \R$.
Based on a sample of i.i.d.\ random variables $V_1, \ldots, V_n \in \R$ from $p$ we seek to estimate this density via maximizing the normalized
log-likelihood function
\bea
    \ell(\varphi) &=& n^{-1} \sum_{i=1}^n \log p(V_i) \ = \ n^{-1} \sum_{i=1}^n \varphi(V_i)
\eea
over all concave functions $\varphi: \R \to [-\infty, \infty)$ such that $\int \exp \varphi(x) \dd x = 1$.

The merits of imposing log-concavity as a shape constraint have been described in detail in \cite{balabdaoui_09}, \cite{cule_09}, \cite{walther_08}, and
\cite{duembgen_logcon10}. The most relevant of those properties in the current context are that many parametric models have log-concave densities.
Examples include the Normal, Exponential, 
Uniform, $\text{Gamma}(r,\lambda)$ for $r\ge 1$, $\text{Beta}(a,b)$ for $a,b \ge 1$, generalized Pareto if the tail
index $\gamma$ is in $[-1, 0]$, Gumbel, Fr\'{e}chet, Logistic or Laplace, to mention only some of these models.
In addition, log-concavity can be considered a straightforward generalization of normality that shares many properties of the Normal distribution \citep{schuhmacher_11},
but is much more flexible.
For these reasons, assuming log-concavity
offers a flexible nonparametric alternative to purely parametric models that includes a wide range of asymmetric, unimodal densities.


The crucial feature of $\hat p_n$, the log-concave density estimator and maximiizer of $\ell(\varphi)$, for our intended application is that the corresponding CDF estimator $\hat P_n$ is,
under some regularity conditions and if the true density is in fact log-concave, asymptotically equivalent to $\mathbb{P}_n$. As a consequence, $\hat P_n$ can be regarded as a smoother of the
empirical distribution function $\mathbb{P}_n$ for finite sample sizes $n$ and this will directly translate into a smooth estimator for the ROC curve.
Apart from the asymptotic result in Theorem~\ref{theo: Fhat} given the appendix, \citet[Corollary 2.5]{duembgen_09} have also derived additional features
of $\hat P_n$ for finite $n$ that further support the plug-in strategy indicated above. Namely, that $\hat P_n(V_1) = 0, \hat P_n(V_n) = 1$, and that $\hat P_n(x)
\in [\mathbb{P}_n(x) - n^{-1}, \mathbb{P}_n(x)]$
whenever $x$ is a knot point of the piecewise linear function $\hat \varphi_n$. 

To conclude the discussion of log-concave density estimation we would like to mention the kernel smoothed version $\hat p^*_n$ of $\hat p_n$ introduced in
\citet[Section 3]{duembgen_09} and generalized in \cite{chen_11}.
Closed formulas can be derived to compute $\hat p^*_n$ and the corresponding CDF $\hat P^*_n$, see \cite{duembgen_logcon10}.
Since log-concavity is preserved under convolution, the smoothed estimate $\hat p^*_n$ remains log-concave if the
applied kernel has this property. In \cite{duembgen_09}, the normal kernel was used with a bandwidth chosen such that the variance of $\hat p^*_n$
equals that of the sample $V_1, \ldots, V_n$. As for $\hat p_n$, this makes $\hat p^*_n$ a fully automatic estimator.
Note that $\hat p_n^*$ is rather close to $\hat p_n$, see \citet[Proposition 3]{chen_11} for a corresponding quantitative result.

\paragraph{Testing for log-concavity.} Typically, shape constraints are
motivated through substantive considerations as above.
However, researchers may want to formally assess the hypothesis of log-concavity. \cite{hazelton_11} adapts Silverman's bandwidth test \citep{silverman_82} to test log-concavity of densities in any
dimension and shows that it has healthy power in several scenarios. The test
works by constructing a kernel estimate of the target density in which
log-concavity is enforced by the use of a sufficiently large bandwidth. The
test statistic is the minimal bandwidth for which log-concavity is achieved
and the null distribution is generated via bootstrap.

In a more exploratory manner, log-concavity can be visually assessed by comparing such an estimate to, e.g., a kernel estimate of the same data.

\begin{figure}[t!]
\begin{center}
\setkeys{Gin}{width=\textwidth}
\includegraphics{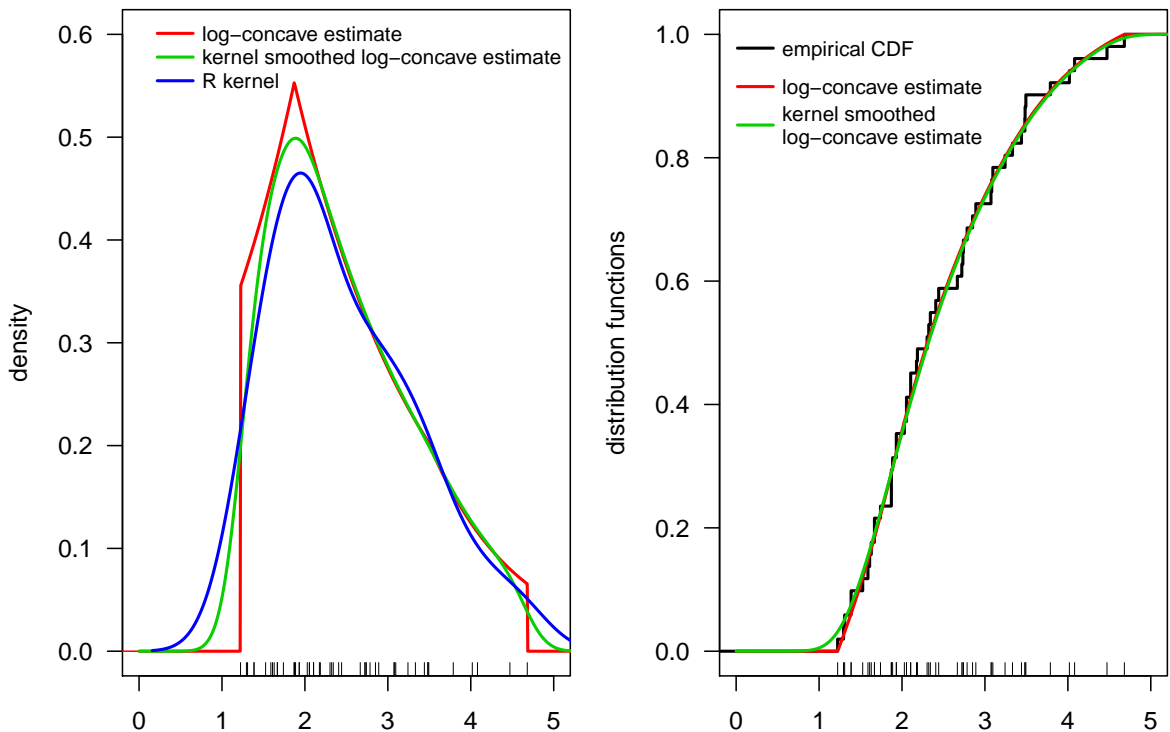}
\end{center}
\vspace*{-0.5cm}
\caption{Pancreas dataset: Log-concave density estimation for log of CA 19.9 measurements, controls only.}
\label{fig: pancreas controls}
\end{figure}

\paragraph{Illustration of the log-concave density estimates.} To illustrate the log-concave estimate, we show in Figure~\ref{fig: pancreas controls}
density estimates applied to the log of the carbohydrate
antigen 19-9 measurements from the pancreatic cancer serum biomarker study, for the pancreatitis patients (i.e. controls) only. This data was
initially analyzed in \cite{wieand_89} and re-assessed, among others, in \citet[Example 1.3.3]{pepe_03}, \citet[Section 6]{cai_04}, \cite{wan_07},
or \cite{zhou_08} to name a few. For details on the dataset we refer to Section~\ref{sec: illustrate roc}. In the
left plot of Figure~\ref{fig: pancreas controls}, the two log-concave density
estimates as well as the standard \proglang{R} kernel estimate are displayed.
It seems safe to assume a log-concave density for this data -- the test by
\cite{hazelton_11} yields a $p$-value of $p = 0.84$ (based on 9999 bootstrap
samples). Note that the log-concave estimate is able to capture the data's
pronounced skewness. The fact that the log of the estimated density is
piecewise linear, see Section~A in the appendix, can be seen through the potential sharp kinks in the
density, as it is the case in Figure~\ref{fig: pancreas controls}. It is
clear that such kinks are alleviated by the smoothed version $\hat p^*_n$.
However, when looking at the CDFs on the right side of Figure~\ref{fig:
pancreas controls}, we see that the seemingly ``nicer'' fit of $\hat p_n^*$ on
the density level does not imply a significant difference for the estimated
CDFs. The right picture further emphasizes that, as claimed above, the
log-concave CDF estimator (whether kernel smoothed or not) can be considered
to be a smoother version of the empirical CDF, a fact theoretically supported
by Theorem~\ref{theo: Fhat}. 

\section{A new ROC curve estimator} \label{newROC}
To define our new ROC curve estimator we first compute log-concave distribution function estimates
$\hat F_m$ and $\hat G_n$ based on the samples $X_1, \ldots, X_m$ and $Y_1, \ldots, Y_n$. The estimates are then plugged into \eqref{def: roc} to get
\bea
    \rocl(t) &:=& R(t; \hat F_m, \hat G_n) \ = \ 1 - \hat G_n(\hat F_m^{-1}(1-t))
\eea for $t \in (0, 1)$. The corresponding estimate based on the CDFs derived from the kernel smoothed log-concave estimates is
denoted by $\rocls(t)$. 
Being a function of the integral of a piecewise exponential function, both estimators are smooth, i.e. they are
at least differentiable on the entire domain $(0, 1)$ and infinitely often differentiable except on the joint set of knot points of
$\hat f_m$ and $\hat g_n$.


In Section~A of the appendix we recapitulate asymptotic results for $\hat p_n$ and $\hat P_n$.
These results can be used to show that $\rocl$, as a process in $t$, is asymptotically equivalent to $\roce$ if the log-concavity assumption holds, see
Section~\ref{main}. This implies that the limiting behavior derived
for $\roce$ in \cite{hsieh_96} in some sense applies to $\rocl$ or that asymptotic confidence intervals derived for $\roce(t)$ are also valid for $\rocl(t)$.
In addition, $\rocl$ is smooth and, as we show in Section~\ref{simulations}, for finite $m$ and $n$ typically more efficient
than $\roce$. We therefore advocate the use of our new estimator as a surrogate for $\roce$ and an alternative to the kernel estimate $\rock$
when it is safe to assume that the constituent densities are log-concave.

It was shown in \cite{lloyd_99} that the estimated ROC curve based on kernels outperforms the empirical, just as $\rocl$. One of the main advantages
of shape constraint estimation in general, and in our current setting in particular, is that such estimates are fully automatic, i.e. they do not necessitate the choice
of a kernel, bandwidth, or some other regularization parameter whose optimal value typically depends on the unknown function to be estimated.

Admittedly, and as it is typical for semiparametric ROC curve estimates, $\rocl$ is generally not invariant with respect to monotone transformations of either $X$
and/or $Y$. However, this is the case for virtually all parametric models.
A ROC curve $R$ (true or estimated) is biased if there exists $p \in (0, 1)$ such that $R(p) < p$.
Log-concave ROC curve estimates can indeed be biased.
However, the bias is in general constrained to regions of $(0, 1)$ that are small, and typically smaller compared to the binormal model.
See \citet[p. 83]{pepe_03} for a detailed discussion of potential bias in binormal models.

Compared to a parametric model with Normal distributions for $F$ and $G$, the log-concave approach is certainly more flexible, at the cost
of only a small reduction in efficiency in case $F$ and $G$ are in fact Gaussian, as we verify in our simulation study in Section~\ref{simulations}.


\section{Main result} \label{main}
In the sequel, a function $g$ is said to belong to the H\"older class ${\cal H}^{\beta, L}(I)$ of functions with exponent $\beta \in (1, 2]$ and
constant $L > 0$ if for all $x, y \in I$ we have $|g'(x) - g'(y)|  \le  L |x - y|^{\beta-1}$.
The claim of asymptotic equivalence of $\rocl$ and $\roce$ under log-concavity is based on Theorem~\ref{theo: asy equi}.
\begin{theorem} \label{theo: asy equi}
Assume that
\bi
\item[C1] $F$ is supported on $I_F = [0, 1]$ with positive density $F' = f$, and $G$ has (potentially infinite) support $I_G \subseteq \R$.
\item[C2] $F$ and $G$ both have log-concave densities $f = \exp \varphi, g = \exp \gamma$ where $\varphi, \gamma$ are both
H\"older-continuous with exponent $\beta \in (1, 2]$ and constant $L > 0$ on $I_F$ and $[A, B] \subset I_G$, respectively.
\item[C3] $F$ and $G$ have log-densities such that $\varphi'(x) - \varphi'(y) \ge C_F(x-y)$ for $0 \le x < y \le 1$ and
$\gamma'(x) - \gamma'(y) \ge C_G(x-y)$ for $A \le x < y \le B$ and two constants $C_F, C_G > 0$.
\item[C4] The sample size $m$ is a function of $n$ such that $n / m \to \lambda > 0$ as $n \to \infty$.
\ei
Then, as $n \to \infty$,
\bea
    \sqrt{n} \sup_{t \in J} \Bl(\rocl(t) - \roce(t)\Br)  &\toP& 0
\eea for $J = [F(A + \delta), F(B - \delta)]$ for every $\delta > 0$.
\end{theorem}

A strong approximation result for the empirical ROC curve process was provided in \citet[Theorem 2.2]{hsieh_96}, see also \citet[Result 5.2]{pepe_03}
or \citet[Section 2]{horvath_08}, and implies the following corollary:
\begin{corollary}\label{cor: lc emp}
Let $\mathbb{B}_1(t)$ and $\mathbb{B}_2(t)$ be two independent Brownian Bridges. Then, as $n \to \infty$,  
\bea
    \sqrt{n}\Bl(\rocl(t) - \roc(t)\Br)&\toD& \mathbb{B}_1\Bl(1-\roc(t)\Br) + \sqrt{\lambda}\frac{g(F^{-1}(1-t))}{f(F^{-1}(1-t))}\mathbb{B}_2(1-t)
\eea uniformly on $J$.
\end{corollary}
Note that the original result in \citet[Theorem 2.2]{hsieh_96} relating the empirical and the true ROC holds a.s. where
Brownian Bridges depending on $n$ are involved. This is the reason why we only get convergence in distribution in Corollary~\ref{cor: lc emp}.
The proof of and some comments on the assumptions of Theorem~\ref{theo: asy equi} can be found in Section~D in the appendix.


\section{Illustration of the new ROC curve estimate} \label{sec: illustrate roc}
\begin{figure}[t!]
\begin{center}
\setkeys{Gin}{width=\textwidth}
\includegraphics{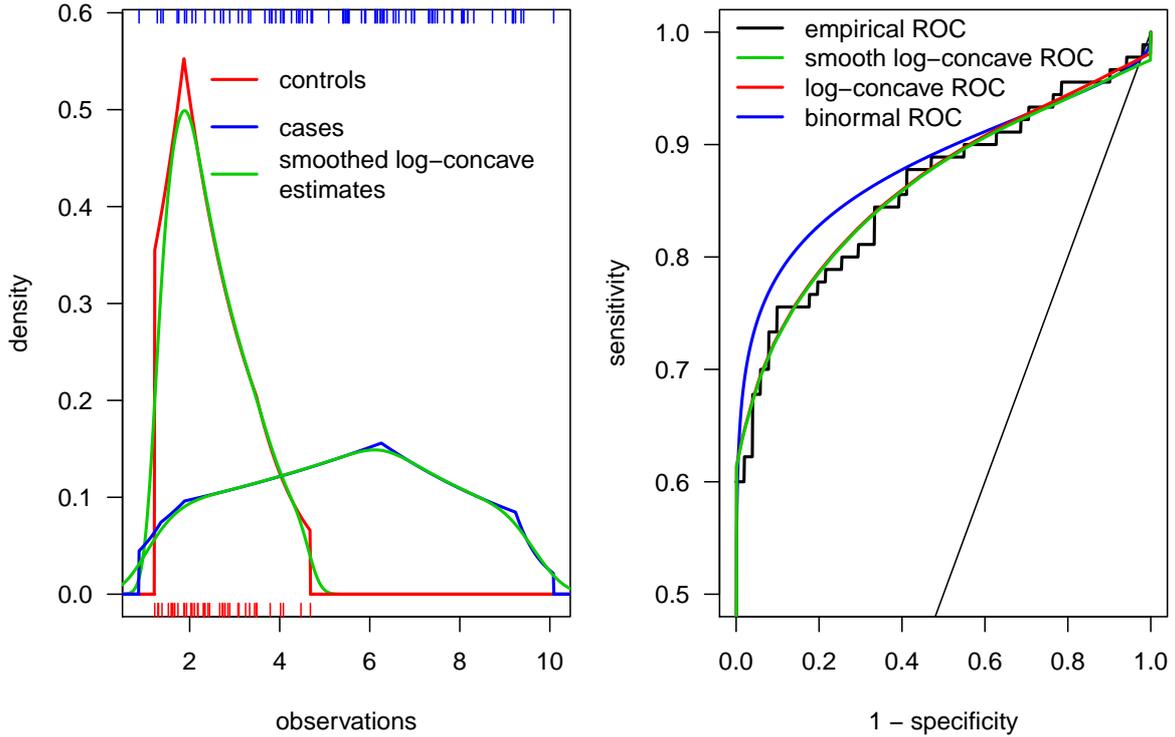}
\end{center}
\vspace*{-0.5cm}
\caption{Pancreas dataset: Log-concave density estimates for log of CA 19.9 measurements and induced ROC curves.}
\label{fig: pancreas ROC}
\end{figure}

We illustrate our new estimator and compare it to the empirical $\roce$ and binormal estimate $\rocn$ on the pancreatic cancer serum biomarker dataset.
The study examined two biomarkers, an antigenic determinant denoted CA125 and a carbohydrate antigen denoted as CA19.9. The data consists of
51 controls that did not have cancer but suffered from pancreatitis and 90 pancreatic cancer
patients. Here, we use the fully parametric binormal model, i.e. we estimate $a$ and $b$ in \eqref{def: binorm roc} directly from the
mean $\mu$ and variance $\sigma^2$ of the underlying distributions.
The results are displayed in Figure~\ref{fig: pancreas ROC} and we observe that (1) Given that the empirical estimate follows the true curve closely,
the binormal model does not seem to be appropriate here, (2) Both log-concave estimates are very close to the empirical estimate, acting as smoothers of
the empirical estimator as discussed above, (3) The difference between the log-concave and the smoothed log-concave estimate in estimation of the
ROC curve is small and mainly concentrated at the boundaries of the domain of the ROC curve. This is not surprising as the corresponding CDF estimates exhibit the same behavior as shown in
Figure~\ref{fig: pancreas controls}. This can be explained by the fact that the discontinuities of $\hat p_n$ at the smallest and largest order statistic of the sample under
consideration are smoothed out by $\hat p_n^*$.
Finally, the test by \cite{hazelton_11} yields a $p$-value to assess the null hypothesis of log-concavity of $p = 0.84$ for the controls and $p = 0.83$
for the cases.

\section{Simulations} \label{simulations}
Having shown asymptotic equivalence to the empirical estimate, it remains to empirically verify that, apart from its inherent properties such as smoothness
and no need to choose any regularization parameters such as kernel and bandwidth,
we additionally enjoy a gain in efficiency by using $\rocl$ instead of $\roce$. To this end, we have performed a simulation study for the scenarios
provided in Table~\ref{tab: scenarios}. Scenarios 1--3 serve as a benchmark for comparing our estimators to the binormal model.
In addition, Scenario 2 has been analyzed in \cite{lloyd_99} and Scenario 3 in \cite{zhou_08}, what enables a direct comparison to these estimators.
Scenario 4 is intended to evaluate the effect of only one distribution being non-normal but skewed. This seems a plausible setup in ROC applications where
values for the controls may be normal but those of cases are right-skewed. Scenario 5 assesses the performance in case both distributions are right-skewed.
Finally, Scenario 6 evaluates the methods for symmetric but non-normal distributions.

\begin{table}[h]
\begin{center}
\caption{Scenarios we simulated from. Ga($\alpha$, $\beta$): Gamma distribution with shape parameter $\alpha$ and rate parameter $\beta$. Log($\mu$, $s$): Logistic distribution with location parameter $\mu$ and shape parameter $s$.}
\label{tab: scenarios}
{\footnotesize
\begin{tabular}{clrlrl}
  \hline
Scenario & $F$ & $m$ & $G$ & $n$ & has been used in \\
  \hline
1 & N(0, 1) & 20 & N(1, 1) & 20 &  \\
  2 & N(0, 1) & 100 & N(1, 1) & 100 & \citet[Figure 1]{lloyd_99} \\
  3 & N(0, 1) & 100 & N(2, 1.2) & 100 & \citet[Figure 1, left plot]{zhou_08} \\
  4 & N(2, 1) & 100 & Ga(2, 1) & 100 &  \\
  5 & Ga(2, 1) & 100 & Ga(4, 1.5) & 100 &  \\
  6 & Log(0, 1) & 100 & Log(2, 1) & 100 &  \\
   \hline
\end{tabular}
}
\end{center}
\end{table}

In our simulations, we compare the empirical ROC estimate $\roce$, the ROC estimates
based on the two variants of log-concave density estimate (maximum likelihood and its smoothed version), the kernel estimate $\rock$
with optimal bandwidths to estimate the ROC curve introduced in \cite{hall_03}, and the fully parametric binormal model.
The latter model is well-specified only if both $F$ and $G$ have a normal density. We would like to emphasize that the binormal model was chosen as a
competitor because it serves as a parametric benchmark if in fact $F$ and $G$ indeed have a normal density and our simulations show, that $\rocl$ and
$\rocls$ are only slightly worse than this benchmark. However, we generally discourage the use of binormal models in practice.

\begin{figure}[p!]
\begin{center}
\setkeys{Gin}{width=\textwidth}
\includegraphics{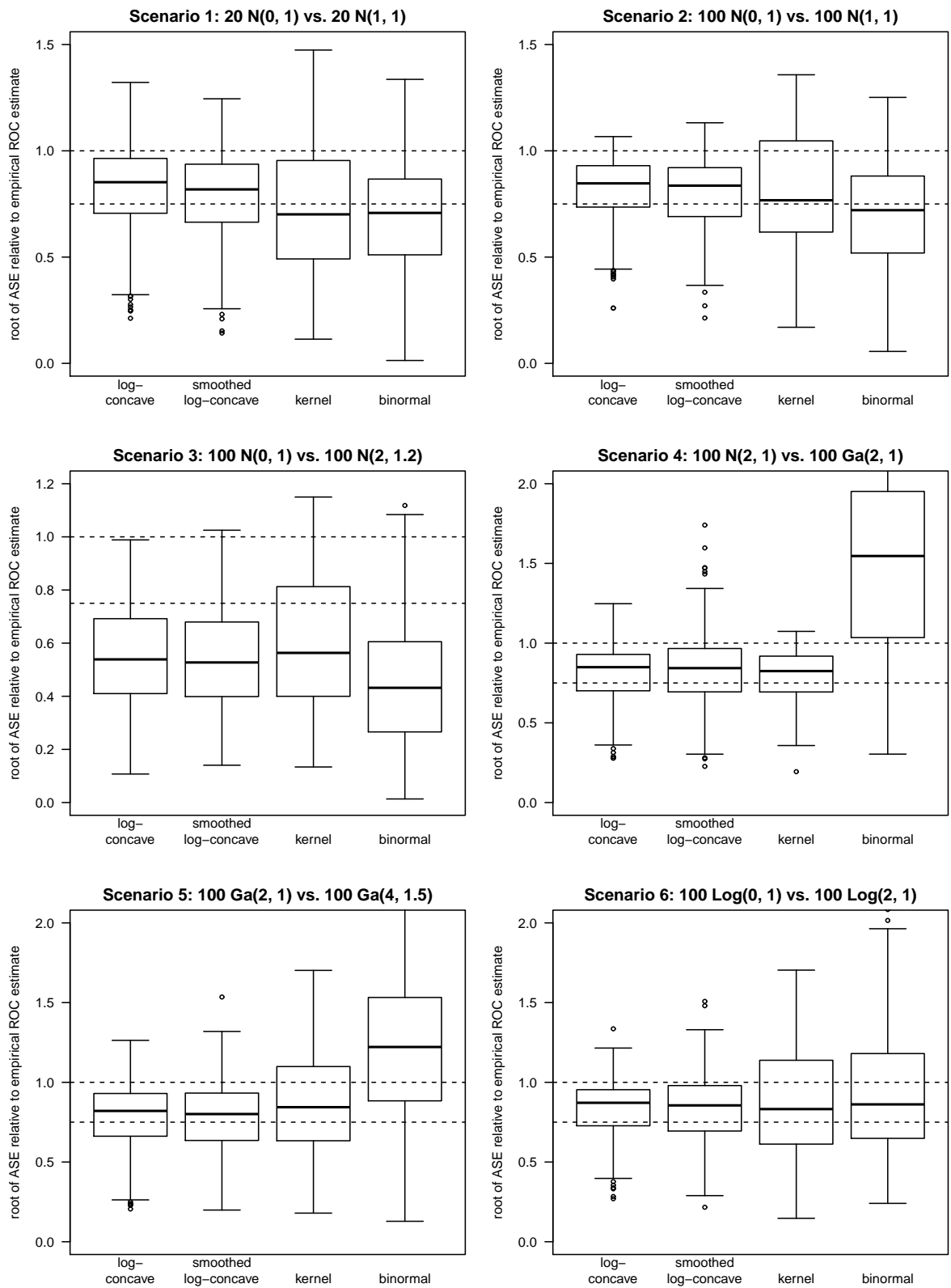}
\end{center}
\vspace*{-0.5cm}
\caption{Results of simulation study. Horizontal dashed lines at 0.75, 1. Note the different scalings of the $y$-axis.}
\label{fig: simul results}
\end{figure}

\begin{figure}
\begin{center}
\setkeys{Gin}{width=0.8\textwidth}
\includegraphics{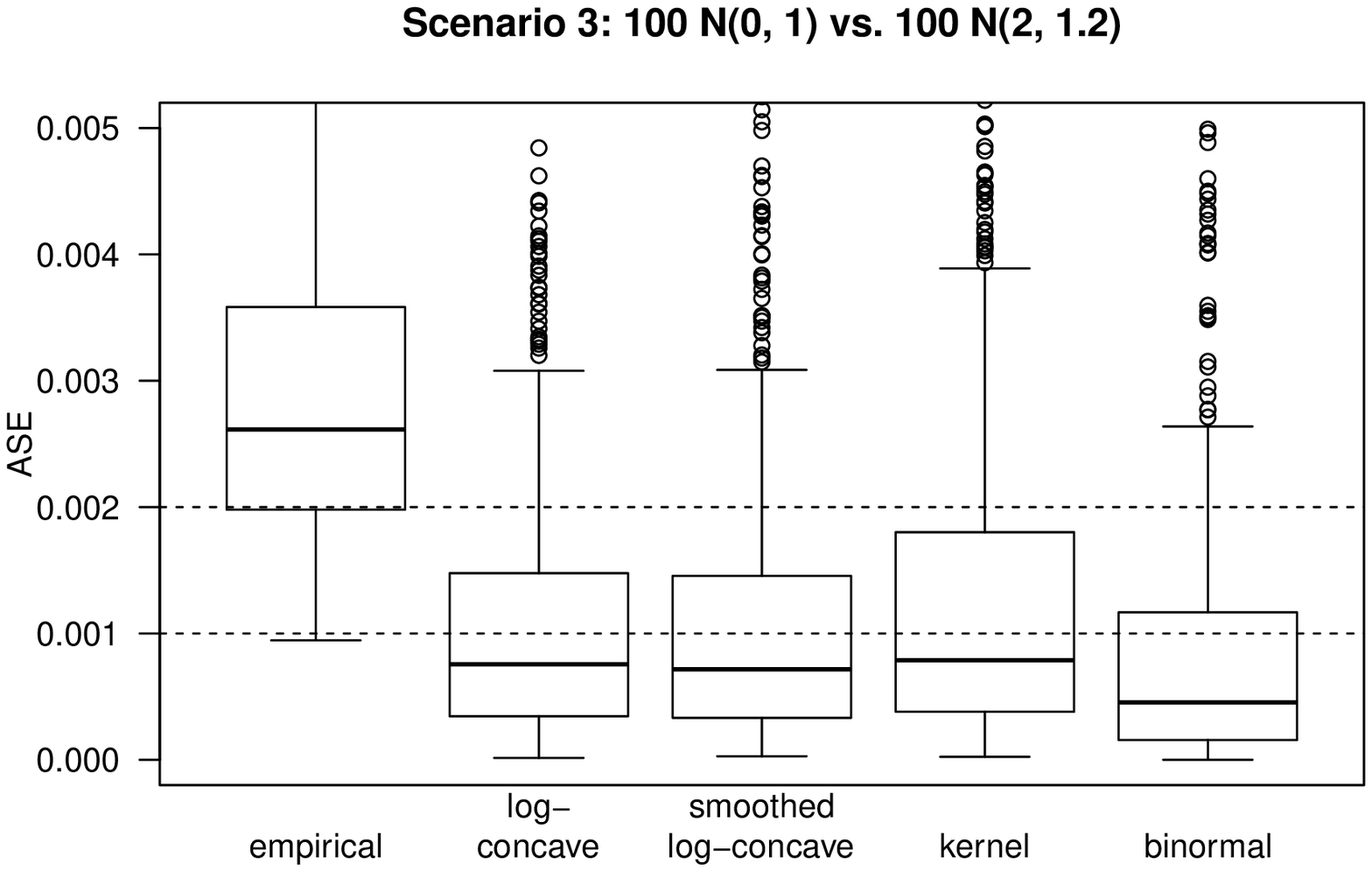}
\end{center}
\vspace*{-0.5cm}
\caption{ASE for Scenario 3, to be compared with \citet[Figure 1, left plot]{zhou_08}. Horizontal dashed lines at 0.001, 0.002.}
\label{fig: simul results s3}
\end{figure}

To evaluate the performance of our competitors we use the average square error ($ASE$), defined for a ROC curve estimate $\hat R$ of a true ROC curve $R$ as
\bea
    ASE(\hat R) &=& n_{\text{grid}}^{-1} \sum_{k=1}^{n_{\text{grid}}} \Bl(\hat R(u_k) - R(u_k)\Br)^2
\eea for grid points $u_i, i = 1, \ldots, n_{\text{grid}}$. This criterion has been used in \cite{zhou_08} to evaluate different variants of binormal
ROC curve estimates. Following the same approach as in the latter paper we choose the $u_i$'s to be equidistant on $[0, 1]$,
$n_{\text{grid}} = 100$ and we generated $M = 500$ samples for each scenario.
Using the same criterion and setting all the simulation parameters to the values already used in \cite[left plot of Figure 1]{zhou_08} enables direct comparison of our
Scenario 3 to their results and we provide the analogous plot in Figure~\ref{fig: simul results s3}. In Figure~\ref{fig: simul results} we display the results for all the six scenarios from Table~\ref{tab: scenarios}, reported as follows:
For each simulation run, the ASE is computed for the four estimators and
$(ASE(\hat R)_j / ASE(\roce)_j)^{1/2}$ for $\hat R \in \{\rocl, \rocls, \rock, \rocn\}$ and $j = 1, \ldots, M$ are reported. Thus, each of the four estimates is benchmarked
with respect to the empirical ROC curve $\roce$. Figure~\ref{fig: simul results} allows for the following observations:
In the purely normal setting (Scenarios 1--3), the log-concave estimators are generally more efficient than the empirical, to an even remarkable extent
for Scenario 3. We attribute this to the fact that here, the ROC curve is rather steep for small values on the abscissa, a shape that $\rocl$ and $\rocls$
are able to more efficiently capture than the empirical.
The kernel estimator $\rock$ outperforms the empirical to an extent comparable to the log-concave estimates.
Finally, compared to the fully parametric binormal model the loss of efficiency for the log-concave versions and the kernel estimate is notably small.

When ``visually averaging'' the pointwise root mean square reductions over $t \in (0, 1)$ in \citet[Figure 1, left plot]{lloyd_99} we see that
their kernel estimate does approximately 15\% better than the empirical estimate, for data simulated from our Scenario 2. This roughly corresponds to our
median reduction in that scenario. 

The gain of the log-concave ROC estimates over the empirical estimate is maintained for non-normal though still log-concave setups (Scenarios 4-6), whereas in these cases the binormal
model is certainly misspecified and therefore underperforming. In Scenarios 5 and 6, $\rocl$ and $\rocls$ provide
values of $(ASE)^{1/2}$ with less variability than $\rock$.

The effect of smoothing the maximum likelihood density estimate in ROC curve estimation is for all the analyzed scenarios rather small.
This is to be expected given that these two estimates are close \citep[Propositon 3]{chen_11} and the fact that they are integrated when computing the
ROC curve.

In \cite{zhou_08} a new estimate for $a$ and $b$ in the semiparametric normal model is introduced and compared to a few competitors, e.g. the one
from \cite{cai_04}. Note that all these estimators are rather complicated to compute and, to the best of our knowledge, no easy accessible
implementation for these estimators is available. The left plot in Figure 1 of \cite{zhou_08} reports results for our Scenario 3 and the analogous
result using the estimators from our simulation study is displayed in Figure~\ref{fig: simul results s3}. Although the estimates
in \citet[Figure 1, left plot]{zhou_08} were explicitly tailored for the binormal model, our log-concave estimates are only slightly, if at all,
less efficient, which is consistent with the results shown in Figure~\ref{fig: simul results}. The same applies to $\rock$.

\section{Bootstrap confidence intervals}\label{confints}
To construct a confidence interval around $\rocl$ at a fixed point $t \in [0, 1]$, one can, by virtue of Theorem~\ref{theo: asy equi}, use standard errors based on asymptotic theory for $\roce$.
Asymptotically, such an interval maintains the pre-specified coverage probability, provided that the log-concavity assumption holds.
However, to exploit the gain in efficiency for finite samples in computation of confidence intervals for values of the true ROC curve,
we suggest to proceed as sketched in \cite{du_09} for a similar spline estimator. Namely, draw $B$ bootstrap resamples $\{X_i^\#\}_{i=1}^m$ and $\{Y_i^\#\}_{i=1}^n$ from the original data and compute for each of these resamples the estimator $\rocl$. This yields bootstrapped versions $\rocl^\#$ of the estimated ROC curve.
The $(1-\alpha)$ confidence interval at a point $t \in [0, 1]$ can then be based on the quantiles of this bootstrap distribution, i.e. we compute
$[(\rocl^\#(t))_{\alpha / 2}, (\rocl^\#(t))_{1-\alpha / 2}]$ where $(\rocl^\#(t))_{\alpha / 2}$ and $(\rocl^\#(t))_{1-\alpha / 2}$
are the $(\alpha/2)$- and $(1-\alpha/2)$-quantile of the $\rocl^\#(t)$'s from the bootstrap samples.

To assess the performance of these intervals we proceeded as follows: For the scenarios in Table~\ref{tab: scenarios} we computed the above intervals at $t \in \{0.1, 0.3, 0.5, 0.7, 0.9\}$ for $B = 500$ and $1-\alpha = 0.95$. As a benchmark we compare the proportion of these intervals out of $M = 500$ simulation runs that cover the true value $\roc(t)$ to the same proportion computed using the BTII interval proposed in \cite{zhou_05}. BTII is a bootstrap confidence interval that does not only estimate the true positive fraction at a fixed $t$, but also accounts for the uncertainty due to the fact that we estimate the quantile corresponding to $t$ based on the controls. \cite{zhou_05} show via simulations that BTII has superior coverage accuracy and shorter interval length compared to other approaches over a wide range of scenarios and for sample sizes comparable to those in our simulation setup.

\begin{figure}[t!]
\begin{center}
\setkeys{Gin}{width=1\textwidth}
\includegraphics{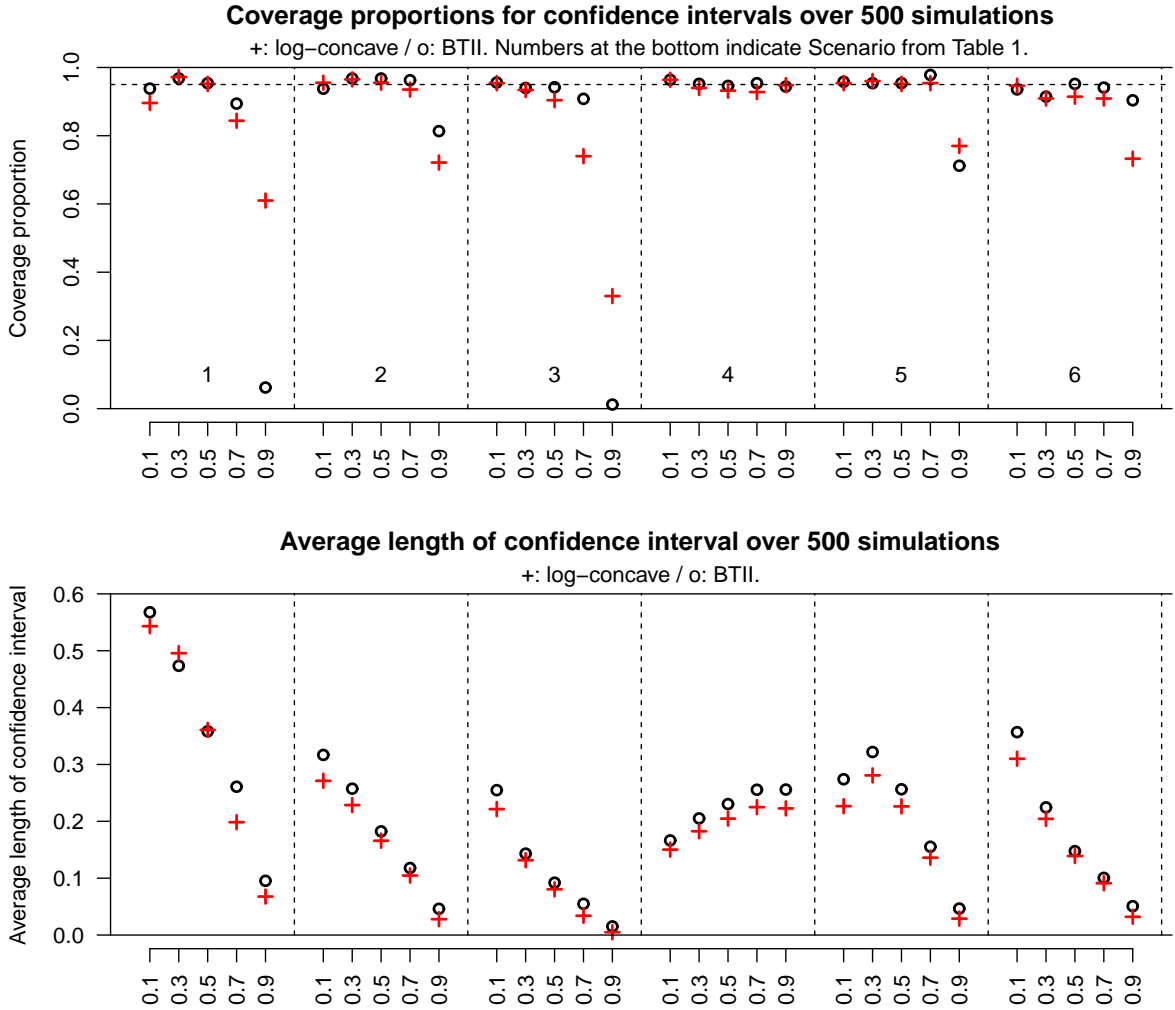}
\end{center}
\vspace*{-0.5cm}
\caption{Performance of confidence intervals for $\roc$ at a fixed $t$. Comparison of bootstrap intervals based on $\rocl$ and the BTII intervals discussed in \cite{zhou_05}.}
\label{fig: simulCI}
\end{figure}

In Figure~\ref{fig: simulCI} we present the results. We find that coverage proportions are basically identical for BTII and the interval based on $\rocl$, for $t \le 0.7$. \cite{zhou_05} noted that the BTII interval performs poorly for values of $t$ that correspond to high ($\ge 0.95$) sensitivities and small sample sizes. Looking at the smallest values of $t$ where the true $\roc(t) \ge 0.95$ in our six scenarios from Table~\ref{tab: scenarios}, we get values of $0.74 / 0.74 / 0.49 / 0.95 / 0.77 / 0.72$, i.e. our results are in line with that rule-of-thumb: for Scenarios 1 and 3, BTII is performing very poorly for $t = 0.9$, whereas the interval based on $\rocl$ has a better performance. Both methods approximately have the same difficulties in reaching the prescribed level for $t = 0.9$ in scenarios 2, 5, 6.
However, as can be inferred from the lower plot in Figure~\ref{fig: simulCI}, the intervals based on $\rocl$ generally yield confidence intervals with a shorter average length over the $M = 500$ simulations, a gain that is in line with the simulation results in Section~\ref{simulations}.

\section{Performance if $f$ and/or $g$ is not log-concave}\label{misspec}
First of all, we would like to point to recent work on estimation of a log-concave density under misspecification. \cite{cule_10_ejs} show that
the log-concave density estimator $\hat f_n$ converges to the density $g$ that minimizes the Kullback-Leibler distance to the true density $f$, in
some exponentially weighted total variation norm. If $f$ is indeed log-concave, this implies a strong type of consistency for the log-concave density
estimator.

Now, it is by no means immediate how this generalizes to our ROC curve
estimate $\rocl$. To assess the robustness of $\rocl$, i.e. scenarios where at least one of the constituent distributions does not have a
log-concave density, we extended the simulations reported in Section~\ref{simulations} to incorporate non-log-concave densities, see Section~C in the appendix.
Additional scenarios are given in Table~\ref{tab: scenarios app}. 
The conclusions from these simulations are that (1) $\rocl$ remains competitive even for moderate deviations from log-concavity for either $f$ and/or $g$,
and that (2) for small samples it even still outperforms the kernel-based estimator by \cite{hall_03}.

\section{Conclusions}\label{conclusions}
We have presented a new approach to estimate ROC curves, which is based on plugging in distribution function estimates received from log-concave
density estimates of the cases and controls data instead of their empirical counterparts.
We propose bootstrap confidence intervals at a fixed $t$ and show asymptotic equivalence between the empirical and the new ROC curve estimate if the log-concavity
assumption holds.
The performance of the new method is assessed via simulations and illustrated on a medical dataset.
In what follows, we recapitulate the reasons why we consider our new approach useful.

In applications, it seems sensible to assume that the true underlying ROC curve $R$ is in fact smooth, thus we expect a better performance
of an estimate by imposing some degree of smoothness. In addition, it is visually appealing to have a smooth estimate rather than a jagged function like
$\roce$.

Many parametric densities are in fact log-concave. As illustrated for the normal distribution in Scenarios 1 to 3 of our simulation study,
not much efficiency is lost when considering log-concavity instead of the actual parametric model if the underlying distributions are indeed Normal.
Besides, assuming log-concavity is much more flexible and robust to misspecification than imposing some parametric model, especially the widely used
binormal model.

In our simulation study we have illustrated that over a wide spectrum of distributions, the estimators $\rocl$ and $\rocls$ perform remarkably better
than the empirical ROC curve for finite sample sizes and have an efficiency that is comparable, and in some scenarios slightly better, than the kernel estimator $\rock$. 
Confidence intervals based on $\rocl$ have comparable coverage probabilities but slightly shorter average interval lengths than the BTII interval from \cite{zhou_05}, if the log-concavity assumption indeed holds.

Taking into account these points, we advocate the use of our estimate since (1) we do not loose much, if anything at all, with respect
to $\rock$ or the binormal model and gain substantially with respect to the empirical ROC curve estimate but (2) are much more robust to misspecification
than the binormal model.
As opposed to kernel estimates, both our estimators $\rocl$ and $\rocls$ are fully automatic.

It was shown that, if the underlying densities are log-concave, $\rocl$ is asymptotically equivalent to $\roce$, so our new estimator can be considered a
``smoother'' of the empirical ROC curve. This ``smoothing property'' of $\hat P_n$ has
given rise to at least two applications where empirical performance of a
method could be substantially improved by using $\hat P_n$ in place of
$\mathbb{P}_n$ as an estimator of the CDF. In \cite{mueller_09} it was
demonstrated that using quantiles based on $\hat P_n^*$ instead of order
statistics reduces estimation variability to a much greater extent in
comparison to the bias introduced in tail index estimation in extreme value
theory. This leads to a substantial reduction in mean squared error.
Similarly, in \cite{duembgen_logcon10} a simulation study demonstrates
increased power in the comparison of two distribution functions when using
the Kolmogorov-Smirnov statistics based on the log-concave instead of the
empirical distribution function estimate.
\section{Software and reproducibility}
Estimation of a univariate log-concave density and many additional related quantities is implemented in the \proglang{R} package \pkg{logcondens} \citep{duembgen_logcon10} available from \proglang{CRAN}. We have added the function \code{logConROC} to this package that gives the ROC curve estimates $\rocl$ and $\rocls$ as well as the corresponding area under the curve (AUC) in just one line of code. The bootstrap confidence intervals from Section~\ref{confints} are implemented as the function \code{confIntBootLogConROC\_t0} in \pkg{logcondens}.
Note that the active-set algorithm used to maximize the log-likelihood function for the log-concave density estimate is remarkably efficient so that it only takes seconds to compute $\rocl$ for rather large sample sizes $m$ and $n$, e.g. of order $m = n = 10^4$. For small to moderate sample sizes computation is immediate. The \code{pancreas} dataset used to illustrate our estimate is also included in \pkg{logcondens} and thus readily accessible.
The code to compute the kernel estimator by \cite{hall_03} is available from the author's webpage. All this enables straightforward reproducibility of the plots and simulations.

In addition, the function \code{smooth.roc} in the \proglang{R} package \pkg{pROC} \citep{robin_11} offers the option to fit a ROC curve based on log-concave density estimates.

This document was created using \code{Sweave} \citep{leisch_02}, \LaTeX \citep{knuth_84, lamport_94}, \proglang{R}
2.14.0 \citep{R} with the \proglang{R} packages
\pkg{logcondens} (\citealp{duembgen_logcon10}, \citealp{logcondens}, Version 2.0.6),
\pkg{xtable} (\citealp{xtable}, Version 1.6-0),
and \pkg{cacheSweave} (\citealp{peng_08}, Version 0.6).

\section*{Appendix}

\appendix

\section{Additional details on log-concave density estimation}\label{webA}
Here, we provide additional details and references on log-concave density estimation.

\cite{duembgen_09} show that the maximizer $\lest$ of $\ell$ is unique, piecewise linear on the interval $[V_{(1)}, V_{(n)}]$ with knots only at
(some of the) observations $V_{(i)}$, and $\lest = - \infty$ elsewhere. Here $V_{(1)} \le V_{(2)} \le \cdots \le V_{(n)}$ are the ordered observations,
and a ``knot'' of $\lest$ is a location where this function changes slope. The MLEs $\lest$, $\hat p_n = \exp \lest$ and $\hat P_n(t) = \int_{-\infty}^t \hat p_n(x) \dd x$ are consistent with certain rates of
convergence, see \cite{duembgen_09} and \cite{balabdaoui_09} as well as Theorem~\ref{theo: Fhat} below.

The crucial feature of the estimate in our context is summarized in Theorem~\ref{theo: Fhat}.

\begin{Theorem}[\citealp[Theorems 4.1 and 4.4]{duembgen_09}] \label{theo: Fhat}
Assume that the log-density $\varphi = \log p \in \Hol{\beta}{L}{T}$ for some exponent $\beta \in [1,2]$, some constant $L > 0$ and a
subinterval $T = [A,B]$ of the interior of the support $\{p>0\}$. Then,
\bea
    \max_{t \in T(n,\beta)} \, |\lest - \varphi|(t) & = & \Op \left( \rho_n^{\beta/(2\beta + 1)} \right),
\eea
where $T(n,\beta) := \bigl[ A+\rho_n^{1/(2\beta + 1)},B-\rho_n^{1/(2\beta + 1)} \bigr]$ and $\rho_n = \log(n) / n$. In particular, this entails that for $\beta \in (1, 2]$
\bea
    \max_{t \in T(n,\beta)} \, \bigl| \hat P_n(t) - \mathbb{P}_n(t) \bigr| &=& \op(n_{}^{-1/2})
\eea where $\mathbb{P}_n$ is the empirical CDF of $V_1, \ldots, V_n$.
\end{Theorem}
Note that the result for $\lest$ remains true when we replace $\lest - \varphi$ by $\hat p_n - p$. Furthermore, it is well-known that the rates of convergence
in Theorem~\ref{theo: Fhat} are optimal, even if $\beta$ was known \citep{hasminski_78}. Thus the log-concave (log-)density estimator adapts to
the unknown smoothness of $p$ in the range $\beta \in (1,2]$.

\section{Estimation of the AUC}\label{webD}
In Section~\ref{simulations} we have shown that $\rocl$ and $\rocls$ (and $\rock$) are valuable approaches to estimate the ROC curve.
In this section, we will show that this holds true also in
estimation of the AUC. AUC is the most widely used \citep[Section 4.3.1]{pepe_03} summary measure for ROC curves and typically
reported in the analysis of a diagnostic test. First, note that Theorem~\ref{theo: asy equi} together with \citet[Theorem 2.3]{hsieh_96} implies that
\bea
    \sqrt{n}(\widehat{AUC} - AUC) &:=& \sqrt{n} \int_J \Bl(\rocl(t) - R(t)\Br) \dd t \\
    &\toD& N(0, \sigma_{\text{AUC}}^2),
\eea so that the AUC based on $\rocl$ and $\roce$ share the same distributional limit (under the assumptions of Theorem~4.1)
as well and asymptotic results for the empirical AUC are valid for the AUC based on the log-concave ROC.
The expression for the asymptotic variance $\sigma_{\text{AUC}}^2$ of the empirical AUC is provided in \citet[Theorem 2.3]{hsieh_96},
see also the discussion in \citet[Section 5.2.5]{pepe_03}.

\section{Estimation of $\rocl$ and AUC under misspecification}\label{webB}
Here, we report on additional simulations when either $f$ and/or $g$ are misspecified, i.e. not log-concave. We have chosen
the Lomax densities as these are an often-used and unbiased model. The $t$-
and the normal mixture densities were selected as they are still unimodal but
have slightly ``too heavy'' tails to still be log-concave.

\begin{table}[h]
\begin{center}
\caption{Scenarios we simulated from to assess performance of the new estimator in misspecified models.}
\label{tab: scenarios app}
{\footnotesize
\begin{tabular}{clrlr}
  \hline
Scenario & $F$ & $m$ & $G$ & $n$ \\
  \hline
7 & Lomax(3, 7) & 20 & Lomax(5, 3) & 20 \\
  8 & Lomax(3, 7) & 100 & Lomax(5, 3) & 100 \\
  9 & t(5, 0) & 20 & t(5, 2) & 20 \\
  10 & t(5, 0) & 100 & t(5, 2) & 100 \\
  11 & N(0, 1) & 20 & 0.75$\cdot$N(2.5, 1) + 0.25$\cdot$N(2.5, 3) & 20 \\
  12 & N(0, 1) & 100 & 0.75$\cdot$N(2.5, 1) + 0.25$\cdot$N(2.5, 3) & 100 \\
   \hline
\end{tabular}
}
\end{center}
\end{table}
\begin{figure}[p!]
\begin{center}
\setkeys{Gin}{width=\textwidth}
\includegraphics{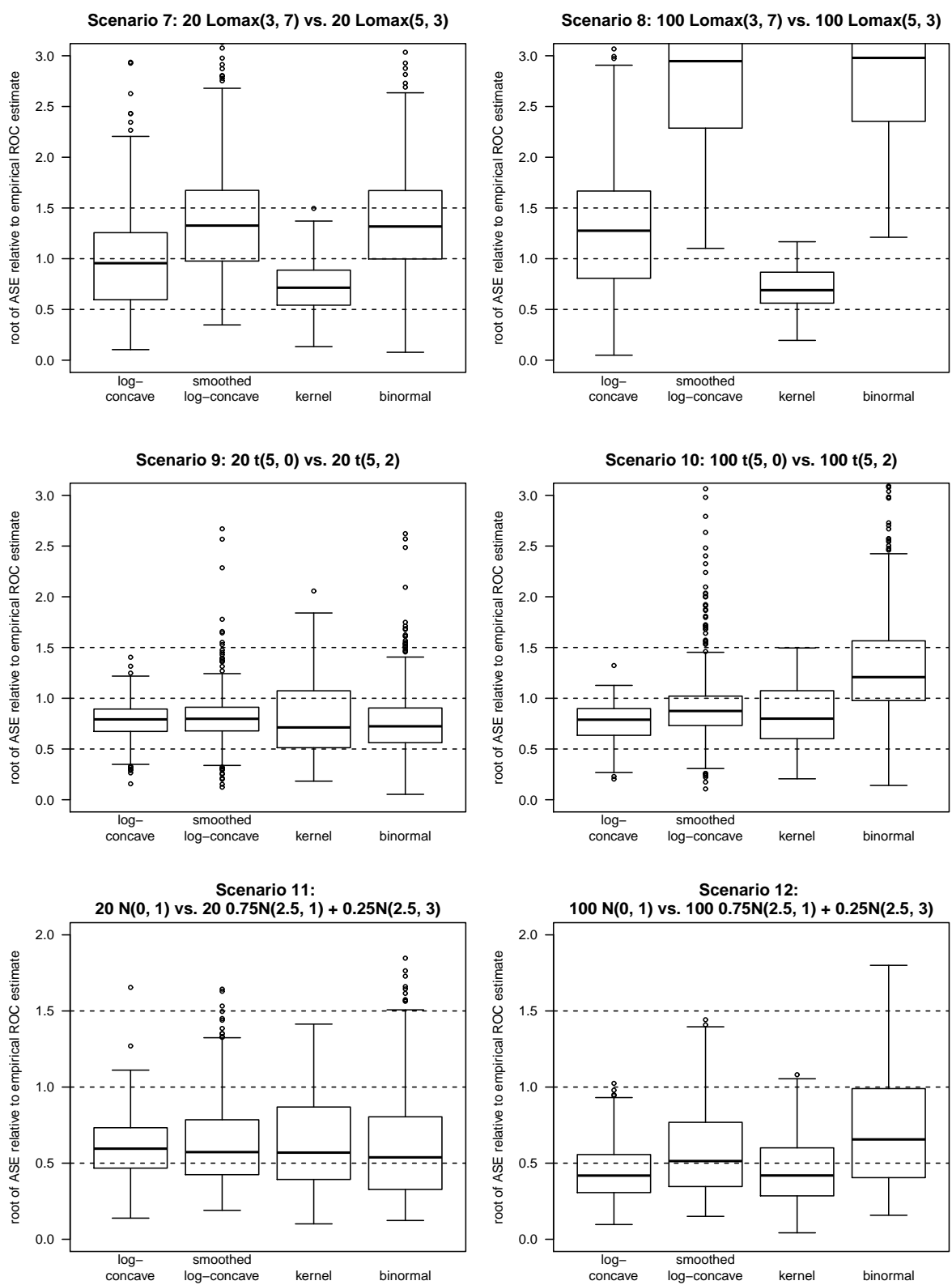}
\end{center}
\vspace*{-0.5cm}
\caption{Results of simulation study for misspecified models. Horizontal dashed lines at 0.5, 1, 1.5. Note the different scalings of the $y$-axis.}
\label{fig: simul results app}
\end{figure}

The ASE relative to the empirical estimate for the simulations from Table~\ref{tab: scenarios app} are
displayed in Figure~\ref{fig: simul results app}. As expected, the estimator by \cite{hall_03} now clearly outperforms $\rocl$. However, the advantage of the latter estimator
is only substantial for the Lomax scenarios whereas for the $t$ and normal mixture setups the two estimators $\rocl$ and $\rock$ perform comparably, even with some advantage for $\rocl$.
We thus conclude that $\rocl$ remains competitive even for moderate deviations from log-concavity in either $F$ and/or $G$, and that for small samples it still
outperforms the kernel-based estimator by \cite{hall_03}.

As a matter of fact, if either $F$ and/or $G$ has a bimodal density it does not seem sensible to use $\rocl$ and we have thus omitted such scenarios from our misspecification simulation study.
It seems safe to assume that in such a scenario, $\rock$ will outperform $\rocl$ and that the latter should not be used based on substantive arguments.

%
%
%
%

As a final remark on misspecified models we would like to emphasize that one typically applies shape constraint estimates based on {\it substantive knowledge}, i.e.
{\it knows} that the densities of $F$ and $G$ are log-concave, or one assesses this claim via formal testing, or via contrasting the log-concave to kernel density estimates.
In addition, as it comes out of the simulations, $\rocl$ is somewhat robust against misspecification.

\section{Proof of Theorem~\ref{theo: asy equi}}\label{webC}

We write
\bea
    \lefteqn{\sqrt{n} \sup_{t \in J} \Bl|\roc(t; \Fm, \Gn) - \roc(t; \hat F_m, \hat G_n)\Br|  \ =} && \\
    &=& \sqrt{n} \sup_{t \in J} \Bl|(1-\Gn(\Fm^{-1}(1-t)))-(1-\hat G_n(\hat F_m^{-1}(1-t)))\Br|  \\
    &=& \sqrt{n} \sup_{t \in J} \Bl|\hat G_n(\hat F_m^{-1}(t)) - \Gn(\Fm^{-1}(t))\Br|  \\ 
    &=& \sqrt{n} \sup_{t \in J} \Bl|\hat G_n(\hat q_{1, t}) - \Gn(\hat q_{1, t}) + \hat g_n\Bl(\hat q_{1, t} + \theta_m(\hat q_{2, t} - \hat q_{1, t})\Br)(\hat q_{2, t} - \hat q_{1, t})\Br|  \\
    &\le& \sqrt{n} \sup_{t \in J} \Bl|\hat G_n(\hat q_{1, t}) - \Gn(\hat q_{1, t})\Br| + \sqrt{n} \sup_{t \in J} \Bl|\hat g_n\Bl(\hat q_{1, t} + \theta_m(\hat q_{2, t} - \hat q_{1, t})\Br)(\hat q_{2, t} - \hat q_{1, t}) \Br| \\
    &=:& T_1(n, m) + T_2(n, m)
\eea
where we defined $\hat q_{1, t} \ = \ \Fm^{-1}(t)$ and $\hat q_{2, t} \ = \ \hat F_m^{-1}(t)$
and $\theta_m$ is some value in $(0, 1)$ for all $m$. By the Bahadur-Kiefer theorem (see e.g. \citealp[Example 3.9.24]{vaart_96} or \citealp[Section 21.2]{vaart_98}) we can write
for any $t \in J$ $\Fm^{-1}(F(t)) = t + (\Fm(t) - F(t)) + \opas(n^{-1/2})$, what together with strong consistency of $\Fm$ \citep[p. 265]{vaart_98} implies that
for any $\varepsilon \in (0, \delta]$ we can find an $n_0$ large enough so that for all $m, n \ge n_0$ almost surely
\bea
    \Bl|\Fm^{-1}(F(A + \delta)) - A - \delta - \rho_n \Br| &\le&\varepsilon
\eea where $\rho_n = \log(n) / n$. But this implies that $\Fm^{-1}( F(A + \delta)) \ge A + \rho_n$ and similarly
$\Fm^{-1}( F(B - \delta)) \le B - \rho_n$ with probability one. Thus,
\bea
    T_1(n, m) & = & \sqrt{n} \sup_{t \in [F(A + \delta), F(B - \delta)]} \Bl|\hat G_n(\hat q_{1, t}) - \Gn(\hat q_{1, t})\Br| \\
    & = &\sqrt{n} \sup_{u \in [\Fm^{-1}\circ F(A + \delta), \Fm^{-1} \circ F(B - \delta)]} \Bl|\hat G_n(u) - \Gn(u)\Br|
\eea and the latter expression is (almost surely) not smaller than
\bea
    \sqrt{n} \sup_{u \in [A + \rho_n, B - \rho_n]} \Bl|\hat G_n(u) - \Gn(u)\Br| &=& \op(1)
\eea by virtue of the second statement of Theorem 4.4 in \cite{duembgen_09}.

As for $T_2(m, n)$ we first note $\hat g_n$ is bounded with probability one (\citealp[Theorem 3.2]{pal_07} and \citealp[Lemma 3]{cule_10_ejs}).
Then, using again \citet[Theorem 4.4]{duembgen_09} one can show that under the assumptions C2 and C3
\bean
    \sup_{t \in J}|\hat F_m(t) - F(t)| &\le& \sup_{t \in J}|\Fm(t) - F(t)| + \op(n^{-1/2}) \label{eq: donsker}
\eean what implies that $\sqrt{n} (\hat F_m - F)$ converges weakly in $D[J]$ (the space of cadlag functions on $J$) to the same limit as $\sqrt{n} (\Fm - F)$, namely to $\mathbb{B} \circ F$
for a standard Brownian Bridge $\mathbb{B}$. But convergence of the estimated distribution function in $D[J]$ implies convergence of the
corresponding quantile process in $D[J]$, by virtue of \citet[Lemma 3.9.23]{vaart_96}. Thus, for some generic constant $C > 0$,
\bea
    T_2(n, m) &=& \sqrt{n} \sup_{t \in J} \Bl|\hat g_n\Bl(\hat q_{1, t} + \theta_m(\hat q_{2, t} - \hat q_{1, t})\Br)(\hat q_{2, t} - \hat q_{1, t}) \Br| \\
    &\le& C \sqrt{\lambda} (1 + \op(1))\sqrt{m} \sup_{t \in J} \Bl|\hat F_m^{-1}(t) - \Fm^{-1}(t)\Br| \\
    &\le& C \sqrt{\lambda}(1 + \op(1))\sqrt{m} \sup_{t \in J} \Bl\{\Bl|\hat F_m^{-1}(t) - F^{-1}(t) + \frac{\mathbb{B} \circ F(F^{-1}(t))}{f(F^{-1}(t))}\Br| + \\
    && \hspace*{0.4cm} \Bl|\Fm^{-1}(t) - F^{-1}(t) + \frac{\mathbb{B} \circ F(F^{-1}(t))}{f(F^{-1}(t))} \Br| \Br\} \ = \ \op(1)
\eea via \citet[Lemma 3.9.23]{vaart_96}. \hfill $\Box$


For the empirical ROC curve this is carried out in \citet[Web supplement]{tang_08} and \citet[Web supplement]{tang_09}.

Now, by Theorem~4.3 in \cite{duembgen_09}, again under the assumptions of Theorem~\ref{theo: asy equi}, we have that $\sqrt{n}(\hat F - F)$ has the same limiting distribution
as $\sqrt{n}(\mathbb{F}-F)$. This in turn implies that the above result carries over to $\rocl$.

\paragraph{Comments on the assumptions of Theorem~\ref{theo: asy equi}.}
As it can be inferred from the proof, the crucial assumption is not log-concavity of the underlying densities, but rather the
Dvoretzky-Kiefer-Wolfowitz inequality \eqref{eq: donsker} which must hold for the CDF estimate. Indeed, log-concavity is only one possible assumption
that entails this property.
The density whose quantile function is involved in the ROC curve estimate needs to have finite support, since it must fulfill the assumptions for
the Bahadur-Kiefer theorem \citep[Section 21.2]{vaart_98}. However, it is easy to see that any truncated log-concave density remains log-concave,
so that this assumption does not seem to be too restrictive.

\bibliographystyle{ims}
\bibliography{C:/rufibach/depot/bibtex/stat}

\def\cprime{$'$}
\begin{thebibliography}{48}
\expandafter\ifx\csname natexlab\endcsname\relax\def\natexlab#1{#1}\fi
\expandafter\ifx\csname url\endcsname\relax
  \def\url#1{\texttt{#1}}\fi
\expandafter\ifx\csname urlprefix\endcsname\relax\def\urlprefix{URL }\fi

\bibitem[{Balabdaoui et~al.(2011)Balabdaoui, Jankowski and
  Rufibach}]{balabdaoui_11}
\textsc{Balabdaoui, F.}, \textsc{Jankowski, H.} and \textsc{Rufibach, K.}
  (2011).
\newblock Maximum likelihood estimation and confidence bands for a discrete
  log-concave distribution.
\newblock \textit{ArXiv e-prints} .
\newline\urlprefix\url{http://arxiv.org/abs/1107.3904}

\bibitem[{Balabdaoui et~al.(2009)Balabdaoui, Rufibach and
  Wellner}]{balabdaoui_09}
\textsc{Balabdaoui, F.}, \textsc{Rufibach, K.} and \textsc{Wellner, J.~A.}
  (2009).
\newblock Limit distribution theory for maximum likelihood estimation of a
  log-concave density.
\newblock \textit{Ann. Statist.} \textbf{37} 1299--1331.

\bibitem[{Cai and Moskowitz(2004)}]{cai_04}
\textsc{Cai, T.} and \textsc{Moskowitz, C.~S.} (2004).
\newblock Semi-parametric estimation of the binormal roc curve for a continuous
  diagnostic test.
\newblock \textit{Biostatistics} \textbf{5} 573--586.

\bibitem[{{Chen} and {Samworth}(2011)}]{chen_11}
\textsc{{Chen}, Y.} and \textsc{{Samworth}, R.~J.} (2011).
\newblock {Smoothed log-concave maximum likelihood estimation with
  applications}.
\newblock \textit{ArXiv e-prints} .

\bibitem[{Cule et~al.(2009)Cule, Gramacy and Samworth}]{cule_09}
\textsc{Cule, M.}, \textsc{Gramacy, R.} and \textsc{Samworth, R.} (2009).
\newblock {LogConcDEAD}: An {R} package for maximum likelihood estimation of a
  multivariate log-concave density.
\newblock \textit{Journal of Statistical Software} \textbf{29}.

\bibitem[{Cule and Samworth(2010)}]{cule_10_ejs}
\textsc{Cule, M.} and \textsc{Samworth, R.} (2010).
\newblock Theoretical properties of the log-concave maximum likelihood
  estimator of a multidimensional density.
\newblock \textit{Electronic J. Stat.} \textbf{4} 254--270.

\bibitem[{Cule et~al.(2010)Cule, Samworth and Stewart}]{cule_08}
\textsc{Cule, M.}, \textsc{Samworth, R.} and \textsc{Stewart, M.} (2010).
\newblock Maximum likelihood estimation of a multidimensional log-concave
  density.
\newblock \textit{J. R. Stat. Soc. Ser. B Stat. Methodol.} \textbf{72}
  545--607.

\bibitem[{Dahl(2009)}]{xtable}
\textsc{Dahl, D.~B.} (2009).
\newblock \textit{xtable: Export tables to LaTeX or HTML}.
\newblock {R} package version 1.5-6.

\bibitem[{Du and Tang(2009)}]{du_09}
\textsc{Du, P.} and \textsc{Tang, L.} (2009).
\newblock Transformation-invariant and nonparametric monotone smooth estimation
  of {ROC} curves.
\newblock \textit{Stat. Med.} \textbf{28} 349--359.

\bibitem[{D{\"u}mbgen et~al.(2010)D{\"u}mbgen, H\"usler and
  Rufibach}]{duembgen_07}
\textsc{D{\"u}mbgen, L.}, \textsc{H\"usler, A.} and \textsc{Rufibach, K.}
  (2010).
\newblock Active set and {E}{M} algorithms for log-concave densities based on
  complete and censored data.
\newblock Tech. rep., University of Bern.
\newblock Available at arXiv:0707.4643.

\bibitem[{D{\"u}mbgen and Rufibach(2009)}]{duembgen_09}
\textsc{D{\"u}mbgen, L.} and \textsc{Rufibach, K.} (2009).
\newblock Maximum likelihood estimation of a log-concave density and its
  distribution function.
\newblock \textit{Bernoulli} \textbf{15} 40--68.

\bibitem[{D{\"u}mbgen and Rufibach(2011)}]{duembgen_logcon10}
\textsc{D{\"u}mbgen, L.} and \textsc{Rufibach, K.} (2011).
\newblock \pkg{logcondens}: Computations related to univariate log-concave
  density estimation.
\newblock \textit{Journal of Statistical Software} \textbf{39} 1--28.

\bibitem[{Hall and Hyndman(2003)}]{hall_03}
\textsc{Hall, P.~G.} and \textsc{Hyndman, R.~J.} (2003).
\newblock Improved methods for bandwidth selection when estimating {ROC}
  curves.
\newblock \textit{Statist. Probab. Lett.} \textbf{64} 181--189.

\bibitem[{Harrell and Davis(1982)}]{harrell_82}
\textsc{Harrell, F.~E.} and \textsc{Davis, C.~E.} (1982).
\newblock A new distribution-free quantile estimator.
\newblock \textit{Biometrika} \textbf{69} 635--640.

\bibitem[{Has{\cprime}minski{\u\i}(1978)}]{hasminski_78}
\textsc{Has{\cprime}minski{\u\i}, R.~Z.} (1978).
\newblock A lower bound for risks of nonparametric density estimates in the
  uniform metric.
\newblock \textit{Teor. Veroyatnost. i Primenen.} \textbf{23} 824--828.

\bibitem[{Hazelton(2011)}]{hazelton_11}
\textsc{Hazelton, M.~L.} (2011).
\newblock Assessing log-concavity of multivariate densities.
\newblock \textit{Statist. Probab. Lett.} \textbf{{81}} {121--125}.

\bibitem[{Horv{\'a}th et~al.(2008)Horv{\'a}th, Horv{\'a}th and
  Zhou}]{horvath_08}
\textsc{Horv{\'a}th, L.}, \textsc{Horv{\'a}th, Z.} and \textsc{Zhou, W.}
  (2008).
\newblock Confidence bands for {ROC} curves.
\newblock \textit{J. Statist. Plann. Inference} \textbf{138} 1894--1904.

\bibitem[{Hsieh and Turnbull(1996)}]{hsieh_96}
\textsc{Hsieh, F.} and \textsc{Turnbull, B.~W.} (1996).
\newblock Nonparametric and semiparametric estimation of the receiver operating
  characteristic curve.
\newblock \textit{Ann. Statist.} \textbf{24} 25--40.

\bibitem[{Knuth(1984)}]{knuth_84}
\textsc{Knuth, D.} (1984).
\newblock Literate programming.
\newblock \textit{Computer Journal} \textbf{27} 97--111.

\bibitem[{Lamport(1994)}]{lamport_94}
\textsc{Lamport, L.} (1994).
\newblock \textit{{\LaTeX}: A Document Preparation System}.
\newblock 2nd ed. Addison-Wesley, Reading, Massachusetts.

\bibitem[{Leisch(2002)}]{leisch_02}
\textsc{Leisch, F.} (2002).
\newblock Dynamic generation of statistical reports using literate data
  analysis.
\newblock In \textit{COMPSTAT 2002 -- Proceedings in Computational Statistics}
  (W.~H\"ardle and B.~R\"onz, eds.). Physica Verlag, Heidelberg.

\bibitem[{Lloyd({1998})}]{lloyd_98_jasa}
\textsc{Lloyd, C.} ({1998}).
\newblock Using smoothed receiver operating characteristic curves to summarize
  and compare diagnostic systems.
\newblock \textit{J. Amer. Statist. Assoc.} \textbf{{93}} {1356--1364}.

\bibitem[{Lloyd(2002)}]{lloyd_02}
\textsc{Lloyd, C.~J.} (2002).
\newblock Estimation of a convex {ROC} curve.
\newblock \textit{Statist. Probab. Lett.} \textbf{59} 99--111.

\bibitem[{Lloyd and Yong(1999)}]{lloyd_99}
\textsc{Lloyd, C.~J.} and \textsc{Yong, Z.} (1999).
\newblock Kernel estimators of the {ROC} curve are better than empirical.
\newblock \textit{Statist. Probab. Lett.} \textbf{44} 221--228.

\bibitem[{M\"uller and Rufibach(2009)}]{mueller_09}
\textsc{M\"uller, S.} and \textsc{Rufibach, K.} (2009).
\newblock Smooth tail index estimation.
\newblock \textit{J. Statist. Comput. Simul.} \textbf{79} 1155--1167.

\bibitem[{Pal et~al.(2007)Pal, Woodroofe and Meyer}]{pal_07}
\textsc{Pal, J.~K.}, \textsc{Woodroofe, M.~B.} and \textsc{Meyer, M.~C.}
  (2007).
\newblock Estimating a polya frequency function.
\newblock In \textit{Complex Datasets and Inverse Problems: Tomography,
  Networks, and Beyond}, vol.~54 of \textit{IMS Lecture Notes-Monograph
  Series}. IMS, 239--249.

\bibitem[{Peng and Zhou(2004)}]{peng_04}
\textsc{Peng, L.} and \textsc{Zhou, X.-H.} (2004).
\newblock Local linear smoothing of receiver operating characteristic ({ROC})
  curves.
\newblock \textit{J. Statist. Plann. Inference} \textbf{118} 129--143.

\bibitem[{Peng({2008})}]{peng_08}
\textsc{Peng, R.~D.} ({2008}).
\newblock {Caching and distributing statistical analyses in R}.
\newblock \textit{Journal of Statistical Software} \textbf{{26}} {1--24}.

\bibitem[{Pepe(2003)}]{pepe_03}
\textsc{Pepe, M.~S.} (2003).
\newblock \textit{The statistical evaluation of medical tests for
  classification and prediction}, vol.~28 of \textit{Oxford Statistical Science
  Series}.
\newblock Oxford University Press, Oxford.

\bibitem[{Qiu and Le(2001)}]{qiu_01}
\textsc{Qiu, P.} and \textsc{Le, C.} (2001).
\newblock R{OC} curve estimation based on local smoothing.
\newblock \textit{J. Statist. Comput. Simulation} \textbf{70} 55--69.

\bibitem[{{R Development Core Team}(2010)}]{R}
\textsc{{R Development Core Team}} (2010).
\newblock \textit{R: A Language and Environment for Statistical Computing}.
\newblock R Foundation for Statistical Computing, Vienna, Austria.
\newblock {ISBN} 3-900051-07-0.

\bibitem[{Robin et~al.(2011)Robin, Turck, Hainard, Tiberti, Lisacek, Sanchez
  and Muller}]{robin_11}
\textsc{Robin, X.}, \textsc{Turck, N.}, \textsc{Hainard, A.}, \textsc{Tiberti,
  N.}, \textsc{Lisacek, F.}, \textsc{Sanchez, J.~C.} and \textsc{Muller, M.}
  (2011).
\newblock {p{R}{O}{C}: an open-source package for {R} and {S}+ to analyze and
  compare {R}{O}{C} curves}.
\newblock \textit{BMC Bioinformatics} \textbf{12} 77.

\bibitem[{Rufibach(2006)}]{rufibach_06}
\textsc{Rufibach, K.} (2006).
\newblock \textit{Log-concave density estimation and bump hunting for I.I.D.
  observations}.
\newblock Ph.D. thesis, Universities of Bern and G\"ottingen.

\bibitem[{Rufibach(2007)}]{rufibach_07}
\textsc{Rufibach, K.} (2007).
\newblock Computing maximum likelihood estimators of a log-concave density
  function.
\newblock \textit{J. Statist. Comp. Sim.} \textbf{77} 561--574.

\bibitem[{Rufibach and D{\"u}mbgen(2011)}]{logcondens}
\textsc{Rufibach, K.} and \textsc{D{\"u}mbgen, L.} (2011).
\newblock \textit{logcondens: Estimate a Log-Concave Probability Density from
  iid Observations}.
\newblock {R} package version 2.0.5.

\bibitem[{Schuhmacher et~al.(2011)Schuhmacher, H{\"u}sler and
  D{\"u}mbgen}]{schuhmacher_11}
\textsc{Schuhmacher, D.}, \textsc{H{\"u}sler, A.} and \textsc{D{\"u}mbgen, L.}
  (2011).
\newblock Multivariate log-concave distributions as a nearly parametric model.
\newblock \textit{Statistics \& Decisions} \textbf{28} 1001--1017.

\bibitem[{Silverman(1982)}]{silverman_82}
\textsc{Silverman, B.~W.} (1982).
\newblock On the estimation of a probability density function by the maximum
  penalized likelihood method.
\newblock \textit{Ann. Statist.} \textbf{10} 795--810.

\bibitem[{Tang et~al.(2008)Tang, Emerson and Zhou}]{tang_08}
\textsc{Tang, L.}, \textsc{Emerson, S.~S.} and \textsc{Zhou, X.-H.} (2008).
\newblock Nonparametric and semiparametric group sequential methods for
  comparing accuracy of diagnostic tests.
\newblock \textit{Biometrics} \textbf{64} 1137--1145.

\bibitem[{Tang and Zhou(2009)}]{tang_09}
\textsc{Tang, L.} and \textsc{Zhou, X.-H.} (2009).
\newblock Semiparametric inferential procedures for comparing multivariate
  {ROC} curves with interaction terms.
\newblock \textit{Statist. Sinica} \textbf{19} 1203--1221.

\bibitem[{van~der Vaart(1998)}]{vaart_98}
\textsc{van~der Vaart, A.~W.} (1998).
\newblock \textit{Asymptotic statistics}, vol.~3 of \textit{Cambridge Series in
  Statistical and Probabilistic Mathematics}.
\newblock Cambridge University Press, Cambridge.

\bibitem[{van~der Vaart and Wellner(1996)}]{vaart_96}
\textsc{van~der Vaart, A.~W.} and \textsc{Wellner, J.~A.} (1996).
\newblock \textit{Weak Convergence and Empirical Processes with Applications in
  Statistics}.
\newblock Springer, New York.

\bibitem[{Walther(2002)}]{walther_02}
\textsc{Walther, G.} (2002).
\newblock Detecting the presence of mixing with multiscale maximum likelihood.
\newblock \textit{J. Amer. Statist. Assoc.} \textbf{97} 508--513.

\bibitem[{Walther(2009)}]{walther_08}
\textsc{Walther, G.} (2009).
\newblock Inference and modeling with log-concave distributions.
\newblock \textit{Statist. Sci.} \textbf{24} 319--327.

\bibitem[{Wan and Zhang({2007})}]{wan_07}
\textsc{Wan, S.} and \textsc{Zhang, B.} ({2007}).
\newblock Smooth semiparametric receiver operating characteristic curves for
  continuous diagnostic tests.
\newblock \textit{Stat. Med.} \textbf{{26}} {2565--2586}.

\bibitem[{Wieand et~al.(1989)Wieand, Gail, James and James}]{wieand_89}
\textsc{Wieand, S.}, \textsc{Gail, M.~H.}, \textsc{James, B.~R.} and
  \textsc{James, K.~L.} (1989).
\newblock A family of nonparametric statistics for comparing diagnostic markers
  with paired or unpaired data.
\newblock \textit{Biometrika} \textbf{76} 585--592.

\bibitem[{Zhou and Harezlak({2002})}]{zhou_harez_02}
\textsc{Zhou, X.} and \textsc{Harezlak, J.} ({2002}).
\newblock {Comparison of bandwidth selection methods for kernel smoothing of
  ROC curves}.
\newblock \textit{Statist. Med.} \textbf{{21}} {2045--2055}.

\bibitem[{Zhou and Lin(2008)}]{zhou_08}
\textsc{Zhou, X.-H.} and \textsc{Lin, H.} (2008).
\newblock Semi-parametric maximum likelihood estimates for {ROC} curves of
  continuous-scales tests.
\newblock \textit{Stat. Med.} \textbf{27} 5271--5290.

\bibitem[{Zhou and Qin(2005)}]{zhou_05}
\textsc{Zhou, X.-H.} and \textsc{Qin, G.} (2005).
\newblock A new confidence interval for the difference between two binomial
  proportions of paired data.
\newblock \textit{J. Statist. Plann. Inference} \textbf{128} 527--542.

\end{thebibliography}

\end{document}